\documentclass[prl,twocolumn,superscriptaddress,floatfix,longbibliography]{revtex4}
\usepackage{enumerate}
\usepackage{
physics}
\usepackage{graphicx}
\usepackage{amsmath}
\usepackage{mathtools}
\usepackage{xcolor}
\usepackage{amsfonts}
\usepackage{amssymb}
\usepackage{units}
\usepackage[normalem]{ulem}

\begin{document}
\pdfoutput=1
\title{Ground-State Preparation via Nonlinear Quantum Dissipation}

\author{Alireza Ataei\footnote{Electronic address: alireza.ataei@math.uu.se}}
\affiliation{Department of Mathematics, Uppsala University, Uppsala, Sweden}

\author{Olle Eriksson}
\affiliation{Department of Physics and Astronomy, Uppsala University, Box 516, 
SE-751 20 Uppsala, Sweden}
\affiliation{Wallenberg Initiative Materials Science,
WISE, Uppsala University, Box 516, 
SE-751 20 Uppsala, Sweden}

\author{Vahid Azimi-Mousolou\footnote{Electronic address: vahid.azimi-mousolou@physics.uu.se}}
\affiliation{Department of Physics and Astronomy, Uppsala University, Box 516, 
SE-751 20 Uppsala, Sweden}

\date{\today}

\begin{abstract}
Ground-state preparation of quantum many-body systems is a fundamental challenge across quantum science, with implications for quantum simulation, quantum computing, and the study of complex quantum matter. The exponential growth of Hilbert space and the complexity of entangled eigenstates motivate the development of new approaches that go beyond conventional strategies based on adiabatic evolution, variational optimization, imaginary-time methods, and engineered dissipative processes. Here, we demonstrate that the recently proposed quantum Landau-Lifshitz-Gilbert (QLLG) dynamics [Phys. Rev. Lett. {\bf 133}, 266704 (2024)] provides an intrinsic nonlinear dissipative mechanism for real-time ground-state preparation. Unlike engineered dissipative schemes, QLLG encodes relaxation directly into the quantum equation of motion while preserving the underlying quantum structure of the evolution. The QLLG evolution selectively suppresses excited-state contributions, driving the system toward the lowest-energy eigenstate within the accessible symmetry sector. We show that the dynamics admits a monotonic energy functional and a stable ground-state fixed point, and that for generic initial states in an $N$-qubit Hilbert space the characteristic convergence time scales linearly with $N$ and inversely with the spectral gap. Numerical simulations of an interacting spin chain confirm our analytical predictions. These results establish nonlinear dissipative QLLG dynamics as a new avenue for quantum state preparation, opening a new route for quantum simulation, computation, and state engineering.
\end{abstract}
\maketitle

\textit{Introduction}--Preparing the ground state of an interacting quantum Hamiltonian is a fundamental challenge in quantum science. Ground states determine the equilibrium properties of quantum materials, molecules, and lattice gauge theories, and their efficient preparation is essential for quantum simulation, quantum chemistry, and quantum optimization. While quantum computers promise substantial advantages for simulating many-body systems, these advantages can only be realized if physically relevant low-energy states can be prepared with scalable quantum resources \cite{Feynman1982,Lloyd1996,Georgescu2014}.

Over the past decades, a variety of approaches and protocols have been developed for preparing quantum ground states. Adiabatic state preparation exploits the quantum adiabatic theorem to continuously transform an easily prepared state into the desired ground state, but its runtime is fundamentally limited by the inverse square of the minimum spectral gap \cite{Farhi2000,AlbashLidar2018}. Variational quantum eigensolvers (VQEs) replace long coherent evolutions with hybrid quantum-classical optimization \cite{Peruzzo2014,McClean2016,Cerezo2021}. However, performance of a VQE depends critically on ansatz design and optimization landscapes, which become increasingly challenging for large or highly entangled systems. Moreover, quantum imaginary-time evolution \cite{Foulkes2001,Vidal2007}, quantum signal processing  \cite{LowChuang2017,LowChuang2019,Gilyen2019}, quantum walks \cite{Aharonov1993,Childs2009,Apers2022}, and related approaches have provided alternative routes to low-energy state preparation with different trade-offs in circuit depth, ancilla requirements, and measurement overhead \cite{Motta2020,LinTong2020,Gilyen2019,Chakraborty2024}.

Despite these advances, existing approaches rely on one of three distinct mechanisms for isolating the ground state: slow adiabatic evolution, classical optimization, or measurement- and ancilla-assisted filtering. A natural question is whether ground-state preparation can instead emerge from a physical dynamical principle that continuously suppresses excited-state components during real-time evolution, without variational optimization, postselection, or explicit imaginary-time propagation.

Dissipation provides precisely such a mechanism for ground-state preparation by turning energetic relaxation into a useful dynamical process. In classical physics, friction naturally drives systems toward minimum-energy configurations. In quantum mechanics, however, dissipation is conventionally associated with decoherence arising from uncontrolled coupling to the environment. This apparent obstacle has motivated the development of engineered dissipative quantum dynamics, where dissipation is carefully designed to stabilize desired quantum states rather than destroy coherence. Early works established dissipative quantum state engineering through reservoir engineering and Lindblad evolution as a powerful framework for preparing many-body quantum states and phases \cite{Ding2024,Diehl2008,Verstraete2009,Zhan2026}.

Building on this idea, recent advances have demonstrated that engineered dissipation can provide efficient routes to quantum ground-state preparation. Lindbladian approaches have enabled ground-state preparation with reduced ancillary resources by constructing dissipative generators for which the target state is a stable stationary state \cite{Ding2024}. More recently, rapid dissipative protocols have shown that appropriately designed open-system dynamics can efficiently prepare ground states of broad classes of noncommuting many-body Hamiltonians \cite{Zhan2026}. Closely related developments exploit engineered thermalization to prepare quantum Gibbs states, where dissipative dynamics drive the system toward the thermal equilibrium state of a target Hamiltonian \cite{Temme2011,Chen2025,Ding2025Gibbs,Rouze2026,Mi2024}. Since the Gibbs state approaches the ground state in the zero-temperature limit, thermalization-based methods provide a complementary dissipative route to preparing low-energy and ground states. Together, these developments establish dissipation as a powerful paradigm for quantum ground-state preparation, offering a fundamentally different route from adiabatic evolution, variational optimization, imaginary-time evolution, and other approaches.

Nevertheless, existing dissipative approaches typically realize relaxation through engineered open-system dynamics, such as reservoir engineering, auxiliary degrees of freedom, or Lindblad generators. A fundamental question remains whether ground-state preparation can instead emerge from an intrinsic nonlinear quantum evolution, in which the relaxation mechanism is encoded directly into the equation of motion while retaining the underlying coherent structure of quantum dynamics.

Quantum Landau--Lifshitz--Gilbert (QLLG) dynamics \cite{Liu2024} was recently introduced as a nonlinear dissipative extension of the Schr\"odinger equation inspired by the Landau--Lifshitz--Gilbert equation of micromagnetics \cite{Landau1935,Gilbert1955}. The resulting dynamics combines unitary evolution with norm-preserving energy dissipation, providing a deterministic relaxation mechanism that continuously lowers the expectation value of the Hamiltonian while maintaining the quantum coherence of the evolving state. Although QLLG was originally proposed as a framework for dissipative quantum dynamics, its dissipative nature and implications for quantum algorithms and state preparation have remained unexplored.

In this Letter, we provide a rigorous proof for  dissipative character of QLLG and show that it naturally constitutes a quantum algorithm for ground-state preparation. We prove analytically that the nonlinear dissipative evolution exponentially suppresses excited-state populations while preserving the ground-state component present in the initial state, leading to deterministic convergence toward the lowest-energy eigenstate. For generic initial states of $N$-qubit systems, we show that the convergence time scales linearly with system size and inversely with the spectral gap. Numerical simulations for interacting Heisenberg spin systems confirms the theoretical predictions and demonstrate efficient preparation of many-body ground states.

\textit{Quantum LLG dynamics and dissipation}--Recently, a quantum analog of the LLG dynamics was proposed in \cite{Liu2024}, where the density matrix $\rho$ evolves according to
\begin{equation}
\dot{\rho} = \frac{i}{\hbar}[\rho,H] + i\kappa[\rho,\dot{\rho}],
\label{QLLG}
\end{equation}
with a damping parameter $\kappa > 0$, and $H$ denoting the Hamiltonian of a many-body spin system.

For Eq.~\eqref{QLLG} to be an analog of the conventional LLG equation, it must describe dissipative magnetization dynamics~\cite{Landau1935, Gilbert1955}, meaning that the energy decreases with time. In Ref.~\cite{azimi2025numerical}, this dissipative behavior was confirmed numerically; however, it can also be established analytically. In particular, the second term on the right-hand side guarantees that the energy expectation value, $\Tr(H\rho(t))$, decreases with time, thereby demonstrating the dissipative nature of the master equation. For a time-independent Hamiltonian, this can be shown as follows. Using Eq.~\eqref{QLLG}, the bound $\Tr(\rho^2)\le 1$ \cite{Liu2024}, together with the Cauchy–Schwarz and triangle inequalities for trace norm ($\|A\|= \sqrt{\Tr(A^{\dagger} A )}$ for an operator $A$), we obtain

\begin{align}
   \sqrt{ \Tr(\dot{\rho}^{\dagger} \dot{\rho}) }\leq \frac{1}{\hbar (1-2\kappa)} \sqrt{\Tr([\rho,H]^{\dagger} [\rho,H])}.
\end{align}
Moreover, Eq.~\eqref{QLLG} yields 
\begin{equation}
\begin{aligned}
&\frac{d\Tr(H\rho)}{dt} = \Tr(H\dot{\rho})
= i\kappa \Tr\big(H[\rho,\dot{\rho}]\big)
\\&= -i\kappa \Tr\big(\dot{\rho}[\rho,H]\big).
\end{aligned}
\end{equation}
Substituting Eq.~\eqref{QLLG} into the above expression and again applying the Cauchy–Schwarz inequality, we obtain
\begin{equation}
\begin{aligned}
&\frac{d\Tr(H\rho)}{dt} 
= \frac{\kappa}{\hbar}
\Tr\big([\rho,H]^2\big)  + \kappa^2 \Tr \big( [\rho,\dot{\rho}] [\rho,H]\big) \\&\leq -\frac{\kappa}{\hbar}
\Tr\big([\rho,H]^{\dagger} [\rho,H]\big) \\&+ \frac{2 \kappa^2}{\hbar (1-2\kappa) } \Tr([\rho,H]^{\dagger} [\rho,H])
\le 0,
\end{aligned}
\end{equation}
for $\kappa < \frac{1}{4}$. The equality holds if and only if $[\rho,H]=0$.
Thus, the energy decreases monotonically until 
$\rho$ commutes with $H$, at which point $\rho$ can be diagonalized in the energy eigenbasis of H. Inspired by this observation, we show below that, for a random pure initial state, the density matrix converges to the ground state of the system, thereby providing a new approach to ground-state preparation and quantum optimization.

\textit{Convergence to the ground state}--For a pure initial state, $\rho_0 = \ket{\psi_0}\bra{\psi_0}$, the QLLG evolution is described by a normalized non-unitary propagation,
\begin{equation}\label{eq:puredensityevolution}
\begin{aligned}
&\rho(t)
=
\frac{
e^{-\frac{i}{\hbar} H_{\mathrm{eff}} t}\,
\rho(0)\,
e^{\frac{i}{\hbar} H_{\mathrm{eff}}^{\dagger} t}
}{
\operatorname{Tr}\!\left(
e^{-\frac{i}{\hbar} H_{\mathrm{eff}} t}\,
\rho(0)\,
e^{\frac{i}{\hbar} H_{\mathrm{eff}}^{\dagger} t}
\right)
},
\\&
H_{\mathrm{eff}}:=\frac{1-i\kappa}{1+\kappa^2}H,
\end{aligned}
\end{equation}
where the effective Hamiltonian $H_{\mathrm{eff}}$ is a non-Hermitian rescaled version of the real Hamiltonian of the system. For derivation of Eq. \eqref{eq:puredensityevolution}, see Appendix A.

Under a finite spectral gap, $\Delta E$ (the energy difference between the ground and first excited states), we demonstrate in the Appendix B that contributions from energy eigenstates other than the ground state are exponentially suppressed, i.e,
\begin{align}
\label{eq:error}
     \|\rho(t)-\Pi_{E_0} \ket{\psi_0} \bra{\psi_0}\Pi^{\ast}_{E_0}\|_1 \simeq \frac{\exp\!\left(-\frac{\kappa \Delta E t}{\hbar(1+\kappa^2)}\right)}{|\bra{\psi_0} \Pi_{E_0} \ket{\psi_0}|^2},
\end{align}
where $\|\,\|_1$ is the trace-norm and $\Pi_{E_0} \ket{\psi_0}$ is the normalized projection of $\ket{\psi_0 }$ in the ground state space of $H$. For a randomly chosen pure initial state, $\ket{\psi_0}$, in a Hilbert space of dimension $2^N$, as shown in the Appendix D, we have
\begin{align}
\label{eq:overlapapprox}
\left|\bra{\psi_0} \Pi_{E_0} \ket{\psi_0}\right|^2 \gtrsim  2^{-N}.
\end{align}
This implies that the overlap of a random initial state with the ground-state subspace is almost always nonzero. When Eq. \eqref{eq:overlapapprox} is combined with Eq.~\eqref{eq:error},
we obtain that QLLG with a random initial pure state always converges to the ground state of the system. Moreover, analytical derivation in the Appendix C shows that the convergence time scales as
\begin{align}
\label{eq:stabletime}
\tau \simeq \frac{\log(2)\hbar(1+\kappa^2)}{\kappa \Delta E}N.
\end{align}
This demonstrates a linear scaling of the convergence time with $N$ and an inverse scaling with respect to the energy gap $\Delta E$.

We conclude our analytical findings with a final remark. As shown in the Appendix B, QLLG dynamics for a given initial pure state, $\ket{\psi_0}$, converges to the lowest-energy eigenspace of the system Hamiltonian with which the initial state has non-zero overlap. This behavior highlights a remarkable and powerful feature of QLLG. Although a random initial pure state leads QLLG dynamics to converge to the ground state, the method is not limited to ground-state preparation. By appropriately choosing the initial state, QLLG can also target higher-energy eigenstates. Specifically, if the initial state is constructed such that the lowest-energy eigenstate with non-zero overlap corresponds to a desired excited state of the Hamiltonian, the dynamics will converge to that excited eigenstate.

\textit{Numerical simulations}--To exemplify the results in Eqs. \eqref{eq:error} and \eqref{eq:stabletime}, we consider the one-dimensional spin chains model described by the following Hamiltonian
\begin{equation}
\label{eq:spinhamiltonian}
H
=
\sum_{i=1}^{N-1}
J\boldsymbol{\sigma}_i \cdot \boldsymbol{\sigma}_{i+1}-
h \sum_{i=1}^{N} \sigma_i^z.
\end{equation}
Here, $J$ is a nearest neighbor Heisenberg exchange, $h$ is the strength of external magnetic field, and $\boldsymbol{\sigma}_i =(\sigma_i^x,\sigma^y_i,\sigma_i^z)$ is the vector Pauli matrices. We compare the exact ground and first excited states with the QLLG simulated ground and first excited states across different values of $h$. In what follows, it is assumed that $J=2$, $\hbar=1$, $\kappa= 0.3$, $N=12$. These values are chosen as model parameters to illustrate the analytical results; the method and theory apply to arbitrary physical systems with general parameter values. Note that for the purpose of this paper, this choice is excellent in describing the inherent physical mechanism of this study, but for a physical, magnetic system one would in most cases expect that the Heisenberg exchange is significantly larger than the applied field strength. 

\begin{figure}[t]
    \centering \includegraphics[width=0.45\textwidth]{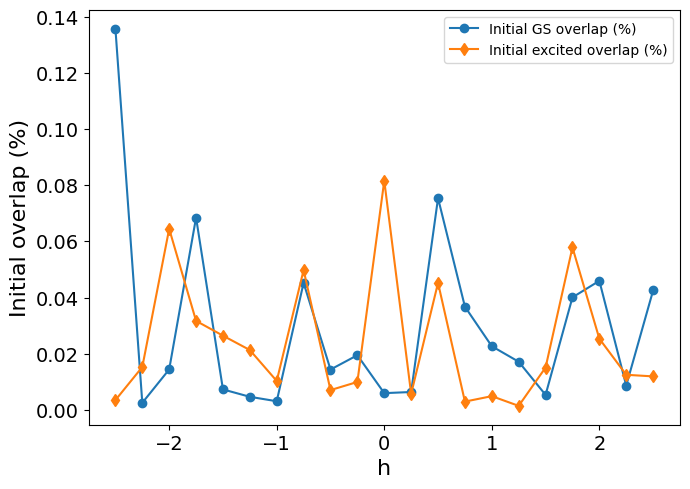}
    \caption{Overlaps of the exact  ground and first excited states with the corresponding initial random states as a function of $h$.}
    \label{fig:initial overlap}
    \vspace{7pt}
\end{figure}

For simulations, we apply the Euler method using 
\begin{equation}
\label{eq:purestateevolutequation}\begin{aligned}
   \partial_t\ket{\psi(t)}    &= -\frac{i (1-i \kappa)}{ \hbar (1+ \kappa^2)}  (H- \bra{\psi(t)} H \ket{\psi(t)}) \ket{\psi(t)}
   \\&= -\frac{i}{\hbar} (H_{\mathrm{eff}}- \bra{\psi(t)} H_{\mathrm{eff}} \ket{\psi(t)}) \ket{\psi(t)},
    \end{aligned}
\end{equation}
deriving in the Appendix A, with small time steps scaled with the same unit as the convergence time in Eq. \eqref{eq:stabletime}. The pure initial states are sampled by drawing the vector coefficients from a complex Gaussian distribution and subsequently normalizing the vector, yielding a random state uniformly distributed on the unit sphere of the Hilbert space. Fig.~ \ref{fig:initial overlap} shows the overlap of the random initial pure states used in the QLLG simulations with the exact ground and first excited states for different values of the magnetic field $h$.

\begin{figure}[t]
\centering
\includegraphics[width=0.45\textwidth]{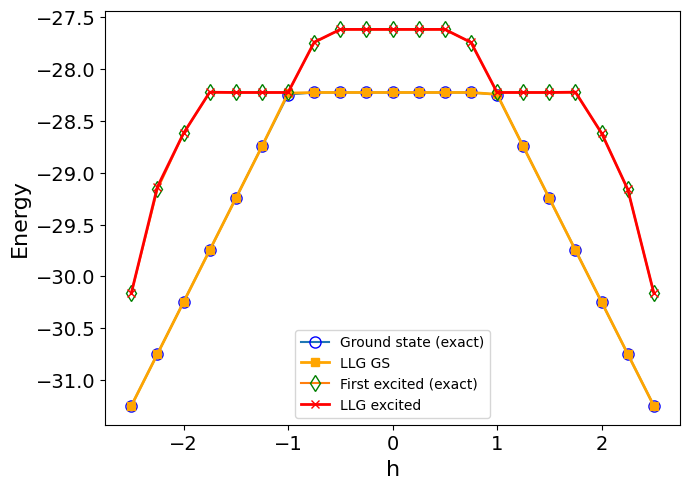}
\caption{Exact and QLLG simulated energies for ground and first excited states.}
\vspace{7pt}
\label{fig: LLG overlap}
\end{figure}

\begin{figure}[t]
\centering
\includegraphics[width=0.45\textwidth]{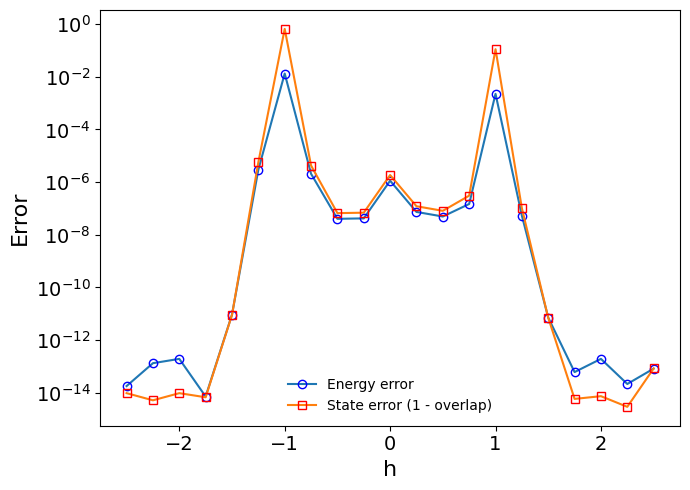}
\caption{The blue curve shows the energy error, defined as the absolute difference between the exact and QLLG-simulated ground-state energies for different values of $h$. The orange curve shows the infidelity of the ground-state overlaps between the exact and QLLG-simulated states as a function of $h$.}
\label{fig:LLG errors}
\vspace{7pt}
\end{figure}

\begin{figure}[t]
    \centering
\includegraphics[width=0.45\textwidth]{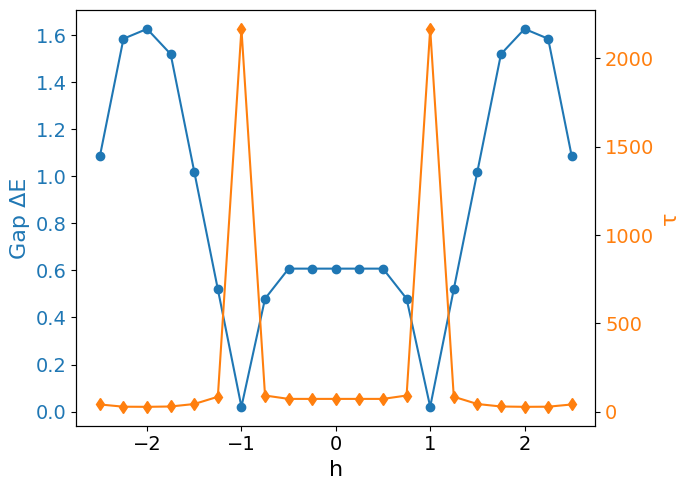}
    \caption{Spectral gap $\Delta E$ (blue curve) and convergence time $\tau$ given in Eq. \eqref{eq:stabletime} (orange curve) as functions of $h$.}  
    \vspace{7pt}
    \label{fig:spectral gap}
\end{figure}

Fig.~\ref{fig: LLG overlap} illustrates the comparison between the exact and QLLG simulated ground and first excited states energies of the Hamiltonian. As shown, energies obtained from the QLLG simulations coincide with the exact results (obtained by exact diagonalization) with a high fidelity for all considered values of $h$. 
Fig.~\ref{fig:LLG errors} shows the energy error rate and the infidelity of the ground states, where the energy error is defined as the absolute difference between the exact and simulated energies, while the infidelity quantifies the deviation of the simulated ground state from the exact ground state, given by
$1 - |\langle \psi^\mathrm{GS}_\mathrm{exact} | \psi^\mathrm{GS}_\mathrm{sim} \rangle|$.
We observe low errors and high fidelity across a broad range of magnetic field values
$h$, except in the vicinity of the critical points with degeneracy between the ground state and first excited state, where the computational complexity/time increases (see Fig.~\ref{fig:spectral gap}). This behavior is consistent with Eq.~\eqref{eq:stabletime}, in which the convergence time is inversely proportional to the energy gap, $\Delta E$. As the gap decreases, the convergence time correspondingly increases. Near the critical points, where the energy gap tends to close, the convergence time therefore grows significantly. This trend is illustrated in Fig.~\ref{fig:spectral gap}, which reflects a trade-off between convergence time and fidelity. Moreover, while algorithmic improvements can enhance performance \cite{azimi2025numerical,mirzaei2025lrei}, the trade-off in Eq.~\eqref{eq:stabletime} remains unavoidable, setting a fundamental limit on the convergence speed. Finally, we notice that despite the occasionally small initial overlaps shown in Fig.~\ref{fig:initial overlap}, the QLLG dynamics consistently converge with high fidelity.

\textit{Conclusion}--We have demonstrated that QLLG dynamics exponentially suppresses excited-state contributions, providing an intrinsic nonlinear dissipative mechanism for real-time ground-state preparation. Our analysis establishes the dynamical stability of the ground state and provides a quantitative real-time convergence law governed by the spectral gap and the initial-state overlap. Together with numerical simulations of interacting qubit systems, these results introduce QLLG as a physically realizable framework for dissipative quantum state preparation and optimization.
Compared to existing approaches based on adiabatic evolution, variational optimization, imaginary-time evolution, or engineered dissipative dynamics, QLLG encodes relaxation directly into the quantum equation of motion through the interplay of coherent precession and nonlinear dissipation. It therefore provides an intrinsic real-time mechanism for ground-state preparation. The predicted real-time convergence is of experimental interest and could be investigated in interacting spin systems with engineered damping, including molecular magnets \cite{kahn2021molecular} and rare-earth atoms on surfaces \cite{pivetta2020measuring}.

More generally, the formulation applies to any quantum model whose dynamics can be mapped onto the QLLG framework, providing a unified real-time approach to ground-state preparation across diverse interacting quantum systems. This opens new opportunities for quantum simulation, quantum optimization, quantum state engineering, quantum chemistry, and the exploration of quantum matter.

\textit{Acknowledgements}--O.E. acknowledges support from the Swedish Research Council (VR), the European Research Council (854843-FASTCORR), the Wallenberg Initiative Materials Science for Sustainability (WISE) funded by the Knut and Alice Wallenberg Foundation (KAW), eSSENCE, STandUP and the Knut and Alice Wallenberg Foundation through grant number 2022.0108.

\section{Appendix}

\subsection*{A. QLLG solution for pure initial states}
Let $\rho_0 =\ket{\psi_0} \bra{\psi_0}$ be a pure initial state and $\rho(t)$ be the solution to QLLG with initial data $\rho_0$. Following the property of the QLLG that the trace and rank of the density matrix are conserved along the evolution, we can write $\rho(t) = \ket{\psi(t)} \bra{\psi(t)}$, where $\braket{\psi(t)}{\psi(t)}=1$. By choosing a proper gauge transformation $\ket{\psi(t)} \to e^{i \gamma(t)} \ket{\psi(t)}$, we can assume that 
\begin{equation}
\label{eq:choicegauge}
    \bra{\psi(t)} (\partial_t \ket{\psi(t)}) = (\partial_t \bra{\psi(t)}) \ket{\psi(t)}=0.
\end{equation}
 Now, using Eq. \eqref{QLLG}, we derive
 \begin{equation}
     \begin{aligned}
&\partial_t(\ket{\psi(t)}) \bra{\psi(t)} + \ket{\psi(t)} \partial_t \bra{\psi(t)} \\&= \frac{i}{\hbar} \left(\ket{\psi(t)} \bra{\psi(t)} H - H \ket{\psi(t)} \bra{\psi(t)}\right) \\&+ i \kappa \left[\rho(t), \partial_t(\ket{\psi(t)}) \bra{\psi(t)} + \ket{\psi(t)} \partial_t \bra{\psi(t)}\right]
\\&= \frac{i}{\hbar} \left(\ket{\psi(t)} \bra{\psi(t)} H - H \ket{\psi(t)} \bra{\psi(t)}\right) \\&+ i\kappa ( \ket{\psi(t)} \partial_t \bra{\psi(t)}-\partial_t(\ket{\psi(t)}) \bra{\psi(t)} ).
     \end{aligned}
 \end{equation}
By multiplying both sides with $\ket{\psi(t)}$ from the right hand-side and using \eqref{eq:choicegauge}, we get 
\begin{equation}
\begin{aligned}
   \partial_t\ket{\psi(t)}    &= -\frac{i (1-i \kappa)}{ \hbar (1+ \kappa^2)}  (H- \bra{\psi(t)} H \ket{\psi(t)}) \ket{\psi(t)}
   \\&= -\frac{i}{\hbar} (H_{\mathrm{eff}}- \bra{\psi(t)} H_{\mathrm{eff}} \ket{\psi(t)}) \ket{\psi(t)}.
    \end{aligned}
\end{equation}
 This concludes the non-Hermitian evolution 
 \begin{equation}
\begin{aligned}
&\rho(t)
=
\frac{
e^{-\frac{i}{\hbar} H_{\mathrm{eff}} t}\,
\rho(0)\,
e^{\frac{i}{\hbar} H_{\mathrm{eff}}^{\dagger} t}
}{
\operatorname{Tr}\!\left(
e^{-\frac{i}{\hbar} H_{\mathrm{eff}} t}\,
\rho(0)\,
e^{\frac{i}{\hbar} H_{\mathrm{eff}}^{\dagger} t}
\right)
},
\\&
H_{\mathrm{eff}}:=\frac{1-i\kappa}{1+\kappa^2}H,
\end{aligned}
\end{equation}
in the main body.

\subsection*{B. Derivation of convergence}
Let $\{ \ket{E_i} \}$ be an eigenbasis of the Hamiltonian $H$, $H\ket{E_i}=E_i\ket{E_i},$ with $E_0$ representing the lowest energy that has a non-zero overlap with an initial state $\ket{\psi_0}$. Consider the following expansion of the initial state
\begin{equation}
\label{eq:expandinitialstate}
\ket{\psi_0} = \sum_i c_i \ket{E_i}.
\end{equation}
Inserting \eqref{eq:expandinitialstate} in \eqref{eq:puredensityevolution}, $\rho(t)$ can be expanded in terms of energy eigenbasis as follows
\begin{align*}
 \rho(t)=\frac{
\sum_{i,j} 
e^{-\frac{\kappa t}{\hbar(1+\kappa^2)}(E_i+E_j)}
e^{\frac{i t}{\hbar(1+\kappa^2)}(E_j-E_i)}
c_i c_j^* \ket{E_i}\bra{E_j}
}{
\sum_i e^{-\frac{2\kappa t}{\hbar(1+\kappa^2)}E_i} |c_i|^2
}.
\end{align*}
By factoring $E_0$ out of exponentials
\begin{equation}
\begin{aligned}
&e^{-\frac{\kappa t}{\hbar(1+\kappa^2)}(E_i+E_j)}
\\&= e^{-\frac{2 \kappa t}{\hbar(1+\kappa^2)} E_0} 
e^{-\frac{\kappa t}{\hbar(1+\kappa^2)} [(E_i-E_0) + (E_j-E_0)]}, \\
&\sum_i e^{-\frac{2\kappa t}{\hbar(1+\kappa^2)}E_i}|c_i|^2 
\\&= e^{-\frac{2\kappa t}{\hbar(1+\kappa^2)} E_0} 
\sum_i e^{-\frac{2 \kappa t}{\hbar(1+\kappa^2)} (E_i-E_0)} |c_i|^2
\end{aligned}
\end{equation}
and defining
\begin{equation}
\gamma := \frac{\kappa}{\hbar(1+\kappa^2)}, \quad 
\omega_{ij} := \frac{E_j-E_i}{\hbar(1+\kappa^2)},
\end{equation}
the QLLG state at time $t$ can be written as
\begin{equation}
\label{eq:energyexpansionfactorization}
\begin{aligned}
\rho(t) = \frac{
\sum_{i,j} e^{-\gamma t [(E_i-E_0)+(E_j-E_0)]} 
e^{i \omega_{ij} t} c_i c_j^* \ket{E_i}\bra{E_j}
}{
p_0 + \sum_{E_i\neq E_0} e^{-2 \gamma t (E_i-E_0)} |c_i|^2
}.
\end{aligned}
\end{equation}
Assume $\Pi_{E_0}$ is the projector onto the energy eigensubspace corresponding to $E_0$, and define the spectral gap and the probability overlap as 
\begin{equation}
\Delta E := \min_{E_i\neq E_0} (E_i-E_0), \,\,
p_0 := |\bra{\psi_0} \Pi_{E_0} \ket{\psi_0}|^2.
\end{equation}
For $i \neq 0$, we have $E_i-E_0 \ge \Delta E$, which implies
\begin{equation}
e^{- \gamma t \big((E_i-E_0) + (E_j-E_0)\big)} \le e^{- 2 \gamma t \Delta E}.
\end{equation}
Thus, as $t \to \infty$, all the terms other than the ones associated with $\ket{E_0}\bra{E_0}$ in Eq. \eqref{eq:energyexpansionfactorization} vanish and
\begin{equation}
\rho(t) = \Pi_{E_0}\ket{\psi_0}\bra{\psi_0}\Pi_{E_0}^{\ast} + 
O\Big(\frac{e^{-\gamma t \Delta E}}{p_0}\Big),
\end{equation}
which is equivalent to 
\begin{align}
     \|\rho(t)-\Pi_{E_0} \ket{\psi_0} \bra{\psi_0}\Pi^{\ast}_{E_0}\|_1 \simeq \frac{\exp\!\left(-\frac{\kappa \Delta E t}{\hbar(1+\kappa^2)}\right)}{|\bra{\psi_0} \Pi_{E_0} \ket{\psi_0}|^2},
\end{align}
in the main body.

\subsection*{C. Error bounds and convergence time}
From Eq. \eqref{eq:energyexpansionfactorization}, we have the trace distance satisfies
\begin{equation}
\|\rho(t) - \Pi_{E_0}\ket{\psi_0}\bra{\psi_0}\Pi_{E_0}^{\ast}\|_1 
\simeq \frac{e^{- \gamma t \Delta E}}{p_0}.
\end{equation}
To achieve an error threshold below $\varepsilon$, we require
\begin{equation}
\frac{e^{- \gamma t \Delta E}}{p_0} \le \varepsilon 
\quad \Rightarrow \quad e^{- \gamma t \Delta E} \le \varepsilon p_0.
\end{equation}
Taking logarithm and substituting $\gamma = \frac{\kappa}{\hbar(1+\kappa^2)}$ yield
\begin{equation}
\begin{aligned}
t &\simeq \frac{\hbar (1+\kappa^2)}{\kappa \Delta E} \Big[ \;
-\log \varepsilon - \log |\bra{\psi_0}\Pi_{E_0}\ket{\psi_0}|^2 \; \Big],
\end{aligned}
\end{equation}
which gives us the convergence time for error bound $\varepsilon$.

\subsection*{D. Typical initial state overlap}
For a random initial state $\ket{\psi_0}$ in a Hilbert space of dimension $D = 2^N$, by Poincar\'e-Maxwell-Borel Lemma,
the overlaps with energy eigenstates 
\begin{equation}
X_i = |\braket{\psi_0}{E_i}|,
\end{equation}
behave approximately as independent, identical random variables, satisfying
\begin{equation}
   \sqrt{D}  X_i \sim \mathcal{N}(0,1),
\end{equation}
for large $N$. Hence, we have the probability
\begin{equation}
\begin{aligned}
&P\left( \frac{\epsilon_1}{D} < \left |\bra{\psi_0}\Pi_{E_0}\ket{\psi_0} \right |^2 < \frac{\epsilon_2}{D}\right) \\&= 2 \int_{\sqrt{\epsilon_1}}^{\sqrt{\epsilon_2}}  \frac{1}{2 \pi} e^{-\frac{z^2}{2}} \, \dd z,
\end{aligned}
\end{equation}
which indicates that the probability of the overlap $\left|\bra{\psi_0}\Pi_{E_0}\ket{\psi_0} \right|^2$ is nonzero and yields \begin{align}
\left|\bra{\psi_0} \Pi_{E_0} \ket{\psi_0}\right|^2 \gtrsim  2^{-N},
\end{align} in the main body if $\epsilon_1 \to 0, \epsilon_2 \to \infty$.

\end{document}